\documentclass[aps,prl,twocolumn,showpacs,superscriptaddress]{revtex4}

\usepackage{amsmath}
\usepackage{amssymb}
\usepackage{graphicx}
\usepackage{textcomp}	%textmu
\usepackage{bbm}

\newcommand{\bra}[1]{\left\langle #1 \right|}
\newcommand{\ket}[1]{\left| #1 \right\rangle}

\begin{document}

\title[]{Bayesian feedback control of a two-atom spin-state in an atom-cavity system}

\author{Stefan~\surname{Brakhane}}\email{brakhane@iap.uni-bonn.de}
\author{Wolfgang~\surname{Alt}}
\author{Tobias~\surname{Kampschulte}}
\author{Miguel~\surname{Martinez-Dorantes}}
\author{Ren\'e~\surname{Reimann}}
\author{Seokchan~\surname{Yoon}}
\affiliation{Institut f{\"u}r Angewandte Physik der Universit{\"a}t Bonn, Wegelerstrasse 8, 53115 Bonn, Germany}
\author{Artur~\surname{Widera}}
\affiliation{Fachbereich Physik der TU~Kaiserslautern und Landesforschungszentrum OPTIMAS, Erwin-Schr{\"o}dinger-Strasse, 67663 Kaiserslautern, Germany}
\author{Dieter~\surname{Meschede}}
\affiliation{Institut f{\"u}r Angewandte Physik der Universit{\"a}t Bonn, Wegelerstrasse 8, 53115 Bonn, Germany}

\date{\today}

\begin{abstract}
We experimentally demonstrate real-time feedback control of the joint spin-state of two neutral Caesium atoms inside a high finesse optical cavity. The quantum states are discriminated by their different cavity transmission levels. A Bayesian update formalism is used to estimate state occupation probabilities as well as transition rates. We stabilize the balanced two-atom mixed state, which is deterministically inaccessible, via feedback control and find very good agreement with Monte-Carlo simulations. On average, the feedback loops achieves near optimal conditions by steering the system to the target state marginally exceeding the time to retrieve information about its state.

\end{abstract}



\maketitle

%\section*{Introduction}
Controlled quantum systems based on neutral atoms or ions and involving multiple particles are providing an experimental platform for the preparation and application of non-classical quantum states~\cite{blatt03}. With such systems, for instance, light can be shed onto non-trivial quantum correlations governing the dynamics of many-body systems~\cite{cirac04}. Tight control over all external (motional) and internal (spin) degrees of freedom is needed to approach unitary evolution of the experimental system, which is the main challenge from the single to the multi particle level.

Passive stabilization of the experimental system is an obvious requirement, e.g. isolation against environmental perturbations. More recently, also active feedback loops are playing an increasing role in stabilizing and preparing quantum states on a time scale longer than the decoherence time~\cite{Sayr11}. Closed loop control is based on a combination of information retrieval (measurements), a controller, and actuators to drive the system towards the desired state. The loop delay time must be short compared to the typical fluctuation time scales of the system. Fast quantum  measurements without strong disturbance of the system, however, typically yield noisy results. Thus it is mandatory to employ real-time methods for the analysis of noisy signals. Experimental advances are aided by the increasing availability of fast and powerful digital signal processors.

A proposal by Balykin and Lethokhov \cite{Baly01} called a feedback method ''information cooling'' for motional control of atoms, emphasizing the close connection of information and control. Feedback control leading to cooling was recently realized \cite{Koch10, Kuba11} with atoms strongly coupled to a high-finesse optical resonator. The method is successful at the single particle level in the strong coupling limit of cavity-QED: The transmission of the optical resonator allows real-time monitoring of the atomic position and motional control via modulation of a trapping potential.

Here, in contrast, we focus on feedback onto the internal atomic state. The projective measurement of the joint internal pseudo-spin state via the cavity transmission level has been proposed for the probabilistic creation of multi-atom entanglement \cite{Sore03}. As a first step towards this goal, we experimentally discriminate the joint discrete two-atom quantum states $\hat{\rho}_{\alpha=0}\!\equiv\! \ket{\downarrow\downarrow}\!\!\bra{\downarrow\downarrow}$, $\hat{\rho}_{\alpha=1}\!\equiv\! 1/2 (\ket{\uparrow\downarrow}\!\!\bra{\uparrow\downarrow} + \ket{\downarrow\uparrow}\!\!\bra{\downarrow\uparrow})$ and  $\hat{\rho}_{\alpha=2}\!\equiv\! \ket{\uparrow\uparrow}\!\!\bra{\uparrow\uparrow}$, where $\alpha$ corresponds to the number of atoms in the spin up state. Our aim is to stabilize the balanced mixed state $\hat{\rho}_{\alpha=1}$ by applying a feedback based on an Bayesian update algorithm~\cite{Reic10}: We assign time-dependent probabilities $p_\alpha(t)$ to the states $\hat{\rho}_\alpha$ and use the measured photon count-rate $n(t)$ of the cavity transmission to determine the conditional probabilities after photon detection according to Bayes' theorem~\cite{Sivia06}. This concept allows in principle to optimally extract the information carried by every measured photon.

\begin{figure}[t]
	\includegraphics{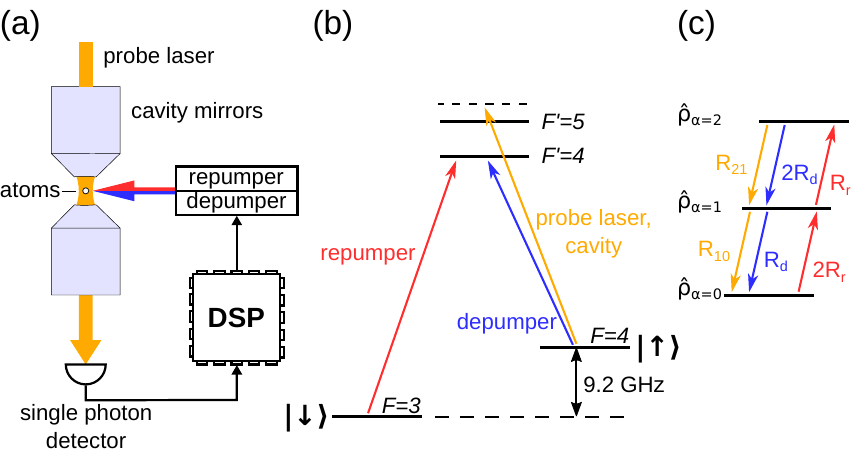}
	\caption{(a) Schematic of experimental setup, (b) Simplified level scheme of Cs, (c) two-atom states and transition rates of probe ($R_{21}$,$R_{10}$), repumping ($R_\text{r}$) and depumping ($R_\text{d}$) laser}
	\label{fig_levelScheme}
\end{figure}

%\section*{Experimental set-up}
In our experiment, the pseudo-spin states are implemented by the two long-lived hyperfine ground states $\ket{F=3} = \ket{\downarrow}$ and $\ket{F=4} = \ket{\uparrow}$ of Caesium. We trap two laser-cooled Cs atoms inside a high finesse optical cavity using a far off-resonant standing wave dipole trap, see Fig.~\ref{fig_levelScheme}(a)\,~\cite{Khuda08,Khuda09}. The cavity resonance frequency is blue detuned from the $F=4 \rightarrow F'=5$ transition of the Cs $\text{D}_2$-line and is on resonance with the frequency of a weak probe laser, see Fig.~\ref{fig_levelScheme}(b).

The detuning of the cavity from the atomic resonance and the position of atoms inside the cavity mode are optimized to achieve long storage times of the atoms and maximally spaced transmission levels: The two-atom states $\hat{\rho}_{\alpha=0,1,2}$ reduce the probe laser transmission by $0\%,\,30\%,\,60\%$ to resolve the different atomic states with the highest contrast~\cite{Reic10}. While the driving of the atom-cavity system by the probe laser allows us to continuously obtain information about the joint atom state via the transmitted light, the probe beam itself also induces spontaneous transitions $\hat{\rho}_{\alpha=2(1)} \rightarrow \hat{\rho}_{\alpha=1(0)}$ by inelastic Raman scattering at rates $R_{21}$ ($R_{10}$), see Fig.~\ref{fig_levelScheme}(c).

The cavity transmission is measured by a single photon counter~\cite{Khuda08} which is connected to a digital signal processor controlling the intensities of both a repumping laser (rate $R_\text{r}$, $F=3 \rightarrow F=4$) and a depumping laser (rate $R_\text{d}$, $F=4 \rightarrow F=3$) in real-time, see Fig.~\ref{fig_levelScheme}.

To exclude atom losses during the experiment, we determine the number of atoms via fluorescence at the beginning and the end of each experimental sequence and we measure the cavity transmission after optical pumping to $\hat{\rho}_{\alpha=2}$ at the end to exclude positioning errors leading to bad coupling.

\begin{figure}
	\includegraphics[width=\linewidth]{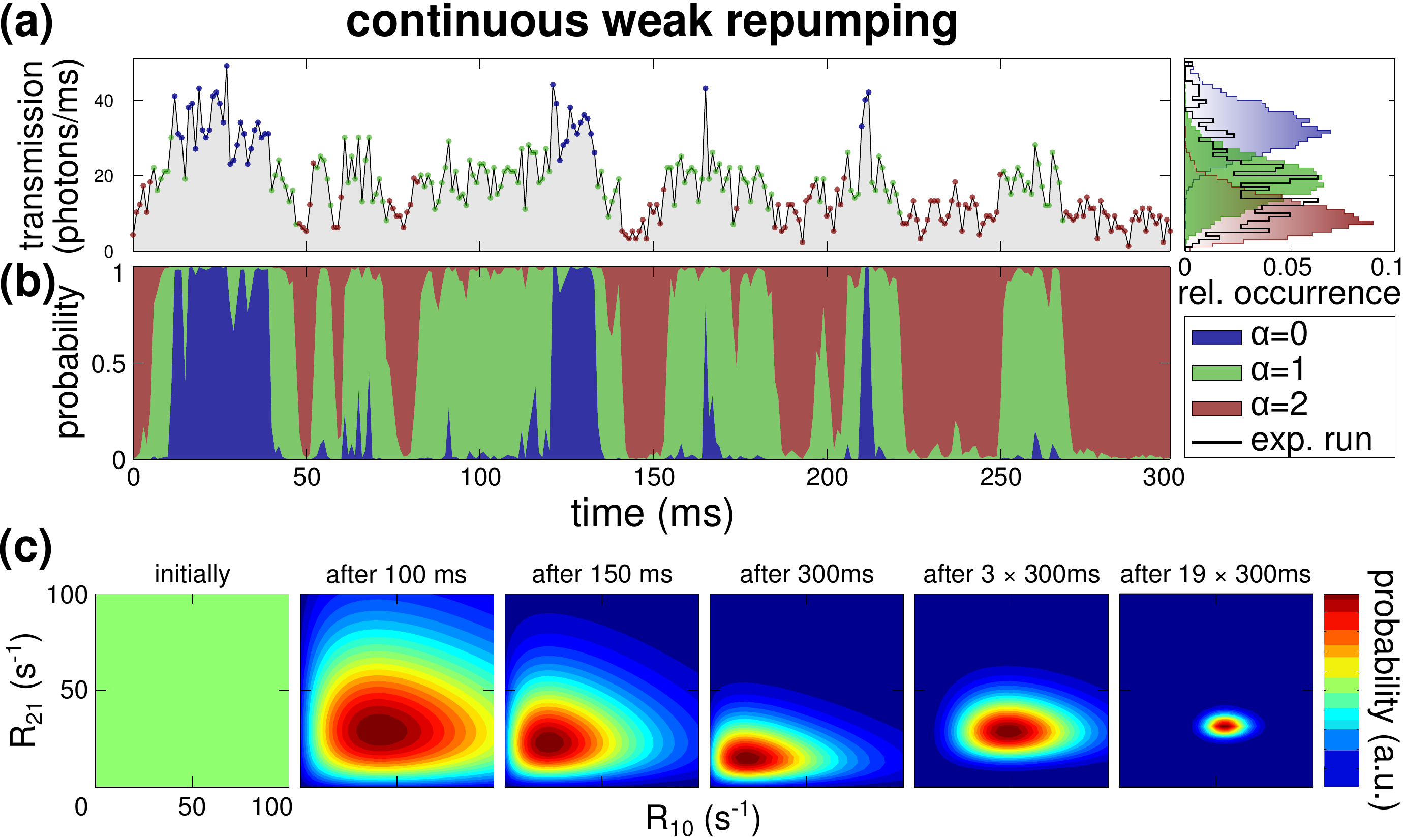}
	\caption{(a) Transmission of the probe laser and photon count histograms, (b) estimated state probabilities for weakly repumped two-atom states for a bin time of $\Delta t = 1\,\text{ms}$, and (c) evolution of the rate probabilities.}
	\label{fig_traceTwoAtoms}
\end{figure}

For the estimation of the rates we choose continuous weak repumping laser intensities ($R_\text{d}=0$, $R_\text{r} \approx R_{10}$), at which we observe abrupt state changes called quantum jumps as shown in Fig.~\ref{fig_traceTwoAtoms}(a)~\cite{Reic10}.

%\section*{Bayesian state estimation}
We estimate the state of the system by assigning occupation probabilities $\mathbf{p}(t) = (p_0(t),p_1(t),p_2(t))^\text{T}$ to the states $\hat{\rho}_\alpha$. At discrete times spaced by the bin time $\Delta t$ we determine the number of transmitted photons $n(t_i)$. Application of Bayes' theorem \cite{Sivia06} yields \textit{a posteriori} state probabilities from \textit{a priori} probabilities $p^\text{pri}_\alpha(t_i)$,
\begin{equation}
    \label{equation_bayesUpdate}
    p^\text{post}_\alpha(t_i) = p(\alpha|n(t_i)) = \frac{p(n(t_i)|\alpha) \, p^\text{pri}_\alpha(t_i) }{\sum\limits_\beta p(n(t_i)|\beta) \, p^\text{pri}_\beta(t_i) } \text{,}
\end{equation}
based on the distribution of conditional probabilities $p(n|\alpha)$ for the same bin time, which are known from separately measured photon count rate histograms for exactly $0,1,2$ atoms coupled to the cavity, see right diagram in Fig.~\ref{fig_traceTwoAtoms}. With no further information available the \textit{a posteriori} probabilities would become the \textit{a priori} probabilities for the following measurement, $\mathbf{p}^\text{post}(t_i)\to \mathbf{p}^\text{pri}(t_{i+1})$ . This procedure can be interpreted as repeated updating of our knowledge about the state of the system based on the measured number of photons $n(t_i)$. The initial \textit{a priori} probabilities $\mathbf{p}^\text{pri}(t_0) = (0,0,1)^\text{T}$ are assigned according to the state $\hat{\rho}_{\alpha=2}$ pepared by optical pumping at the beginning of each experimental sequence.

The average evolution of our system is described by rate equations. Thus we can further improve our know\-ledge of the \textit{a priori} system state at time $t_{i+1}$ by taking into account the evolution from the previous \textit{a posteriori} system state $\mathbf{p}^\text{post}(t_i)$. For weak continuous repumping (rates $\{R_j\}=\{R_{21},R_{10},R_\text{r}\} \ll \Delta t^{-1}$, Fig.~\ref{fig_levelScheme}(c)) the linearized solution is
\begin{equation}
	\label{equation_rateEquation}
	\mathbf{p}^\text{pri}(t_{i+1})\\
	 = \left[\mathbbm{1} + \Delta t
	\begin{pmatrix}
		-2 R_\text{r} & R_{10} & 0 \\
		2 R_\text{r} & -R_{10}-R_\text{r} & R_{21}\\
		0 & R_\text{r} & -R_{21}
	\end{pmatrix}
	\right]
	\cdot
	\mathbf{p}^\text{post}(t_i)
\end{equation}
where $\mathbbm{1}$ is the 3x3 identity matrix.

The model implemented by Eq.~(\ref{equation_rateEquation}) allows to extend Bayes' theorem to generalized probabilities $p(\alpha,\{R_j\})$ for states \textit{and} rates, overcoming the need to determine $\{R_j\}$ by an independent measurement:
\begin{equation}
    \label{equation_bayesRateUpdate}
    p^\text{post}(\alpha,\{R_j\}|n(t)) \propto p(n(t)|\alpha) \, p^\text{pri}(\alpha,\{R_j\})
    \text{.}
\end{equation}
The photon count histograms $p(n|\alpha)$ are not directly affected by the rates if $\{R_j\}\ll \Delta t^{-1}$. Nevertheless, every measurement $n(t_i)$ provides information about the rates since Eq.~\ref{equation_rateEquation} updates our knowledge by predicting an \textit{a priori} distribution for the genera\-lized probabilities $p^\text{pri}(\alpha,\{R_j\})$.

We take an initally flat probability distribution (no know\-ledge) for the rates and evaluate the probabilities on a discrete grid in the four dimensional space of states and rates for each measurement $n(t_i)$. The probability values for any rates or states alone can be calculated using the marginalization rule, e.g.
\begin{equation}
    \label{equation_rateSummation}
    p^\text{post}(\alpha|n(t)) = \sum\limits_{\{R_j\}} p^\text{post}(\alpha,\{R_j\}|n(t))\quad\text{.}
\end{equation}
An example of the time evolution of a free running, weakly repumped system is given in Fig.~\ref{fig_traceTwoAtoms} (b) for the state probabilities $\rho_\alpha$ and in (c) for the distribution of rates $R_{10}$ and $R_{21}$ where, with increasing data accumulation (information gain), a narrow peak emerges. The transition rates and errors are extracted as the expectation values and the rms values of the probability distribution. We stop data acquisition when the uncertainty of the transition rates is $\approx 10\%$ yielding rates $R_{10} {=} (50 {\pm} 6)\,\text{s}^{-1}$, $R_{21}  {=} (35 {\pm} 4)\,\text{s}^{-1}$ and $R_\text{r} {=} (59 {\pm} 5)\,\text{s}^{-1}$, where $5.1\,\text{s}$ of data acquisition corresponding to $\sim 250$ quantum jumps were used as an improvement with a significantly lower number of quantum jumps compared to \cite{Reic10}.

Theoretically, the Bayesian data analysis is independent of the choice of bin time for $\Delta t {\leqslant} R^{-1}_i$ for shot noise limited signals. Experimentally, the analysis yields constant rates in the range of $\Delta t {=} 0.3\,\text{ms}\ldots10\,\text{ms}$. Below $\Delta t{=}0.3\,\text{ms}$ we observe an increase of the extracted transition rates due to a super-poissonian broadening of the photon count histograms $p(n(\Delta t)|\alpha)$~\cite{Reic10}. We attribute this broadening to atom-cavity coupling fluctuations induced by atomic motion. They cause correlations not accounted for by the Bayesian state estimation, which leads to noise affecting the rate estimation. They are more relevant at short bin times where the photon number distributions $p(n(\Delta t)|\alpha)$ are not well separated. We have thus chosen a bin time of $\Delta t = 1\,\text{ms}$ which maintains high time resolution while providing acceptable separation of the photon count histograms. Before closing the feedback loop we measure the photon count histograms and determine the rates $R_{21}$,$R_{10}$ according to Eq.~(\ref{equation_bayesRateUpdate}).

%\section*{Feedback}
\begin{figure}
	\includegraphics[width=\linewidth]{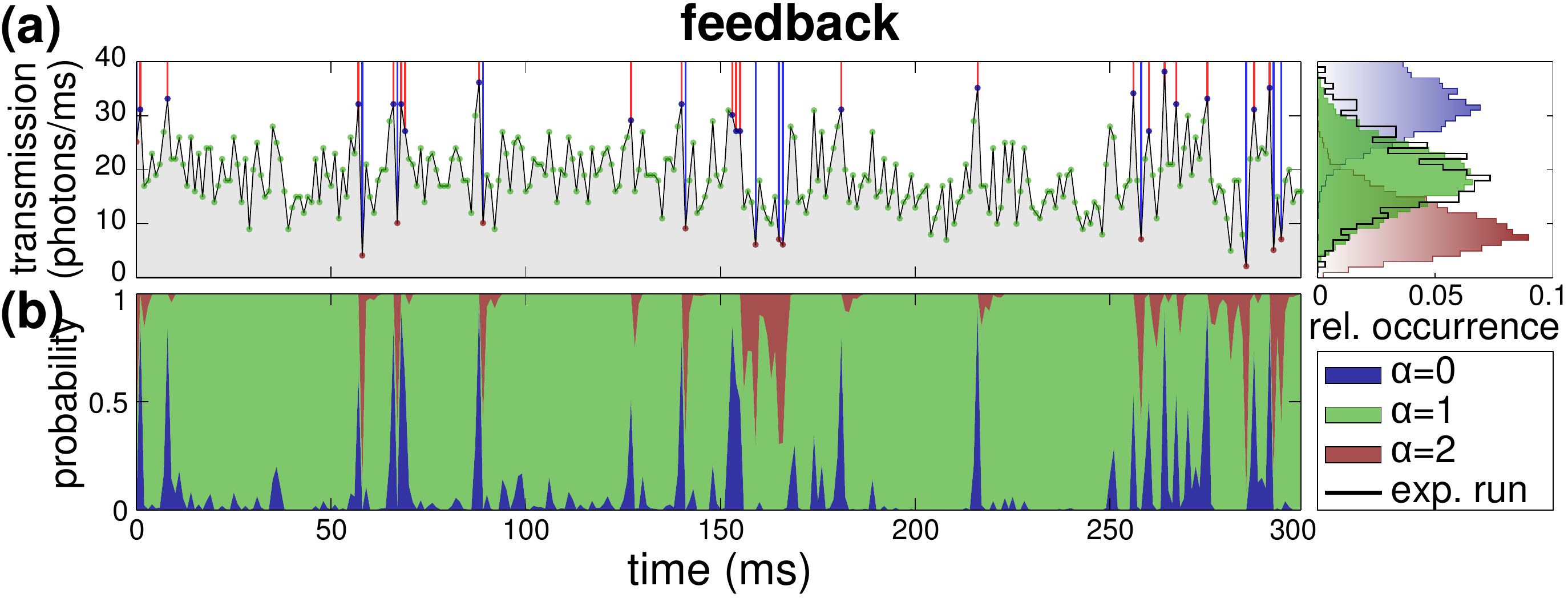}
	\caption{(a) Transmission of the probe laser and transmission histograms of 0 (blue), 1 (green) and 2 (red) coupled atoms and current transmission signal (black). The red and blue lines indicate the incidence of repumping and depumping feedback pulses, respectively. Note that the histogram of the current trace resembles the histogram of one coupled atom to a very high degree. (b) Estimated state probabilities for feedback onto the two-atom state $\hat{\rho}_{\alpha=1}$. A bin time of $\Delta t = 1\,\text{ms}$ has been used for all subfigures.}
	\label{fig_traceFeedback}
\end{figure}

In order to steer the system towards the target state $\hat{\rho}_{\alpha=1}$ we continuously monitor the cavity transmission and use Eqs.~(\ref{equation_bayesUpdate}) and (\ref{equation_rateEquation}) for real-time state estimation. The algorithm controlling the application of short, intense pulses of repumping and depumping laser light minimizes the Kolmogorov distance $k(\mathbf{p}_\text{target},\mathbf{p}(t)) = \frac{1}{2} \sum_\alpha \left| \mathbf{p}_\alpha^\text{target} {-} \mathbf{p}_\alpha(t) \right|$ which quantitatively measures the difference of the estimated time-dependent probabilities from the target state.

The short laser pulses drive state changes with transition probabilities $T_\text{r,d}$ during a single pulse of length $\delta t$. This knowledge is included in our algorithm by multiplying $\mathbf{p}^\text{post}(t)$ with a matrix
\begin{eqnarray}
	\mathbf{M_\text{r}}	
	 & = &
	\begin{pmatrix}
		(1-T_\text{r})^2 & 0 & 0\\
		2T_\text{r}(1-T_\text{r}) & 1-T_\text{r} & 0\\
		T_\text{r}^2 & T_\text{r} & 1
	\end{pmatrix}
\end{eqnarray}
for a repumping laser pulse and accordingly $\mathbf{M_\text{d}}$ for a depumping pulse. The optimal transition probabilities can then be calculated by minimizing the Kolmogorov distance $k(\mathbf{p}_\text{target},\mathbf{M}_i\mathbf{p}^\text{post}(t))$ with respect to $T_i$. Here, the optimal $T_i$ depend on $\mathbf{p}(t)$ and lie within a range of $[0.25,0.5]$ per pulse. However, simulations show that the mean occupation of the target state does not change significantly if the algorithm is simplified as follows: We use a fixed value of $T_\text{r} (T_\text{d})$ and apply a repumping (depumping) laser pulse if $p_0(t)>p_1(t),p_2(t)$ $(p_2(t)>p_0(t),p_1(t))$, respectively. Since this feedback method is less demanding to be technically implemented, we have experimentally set $T_\text{r,d}$ to values that maximize the estimated probability of the target state by fixing the length of the pulses to $\delta t \approx 1.5\,\text{\textmu s}$ and optimizing its intensity. The computation of the closed feedback loop algorithm takes a time of about $6\,\text{\textmu s}$ on our digital signal processor (TMS6713 by Texas Instruments) and can thus be neglected with respect to the update frequency of $1\,\text{ms}^{-1}$.

A typical measurement of the probe transmission with feedback is plotted in Fig.~\ref{fig_traceFeedback}. The red and blue vertical lines in the background of the transmission data indicate a repumping (red) and depumping (blue) laser pulse. The random telegraph pattern of the quantum jumps is strongly suppressed and the state probabilities are dominated by $p_1$. Furthermore, the photon-count histogram of the experimental feedback trace (black) is almost identical with the photon count histogram of a single coupled atom ($\alpha=1$, green), confirming the reliability of our state estimation and feedback scheme. In principle an out-of-loop measurement of the atomic states could be performed via a push-out technique \cite{Schra04}.

\begin{figure}
	\includegraphics[width=1\linewidth]{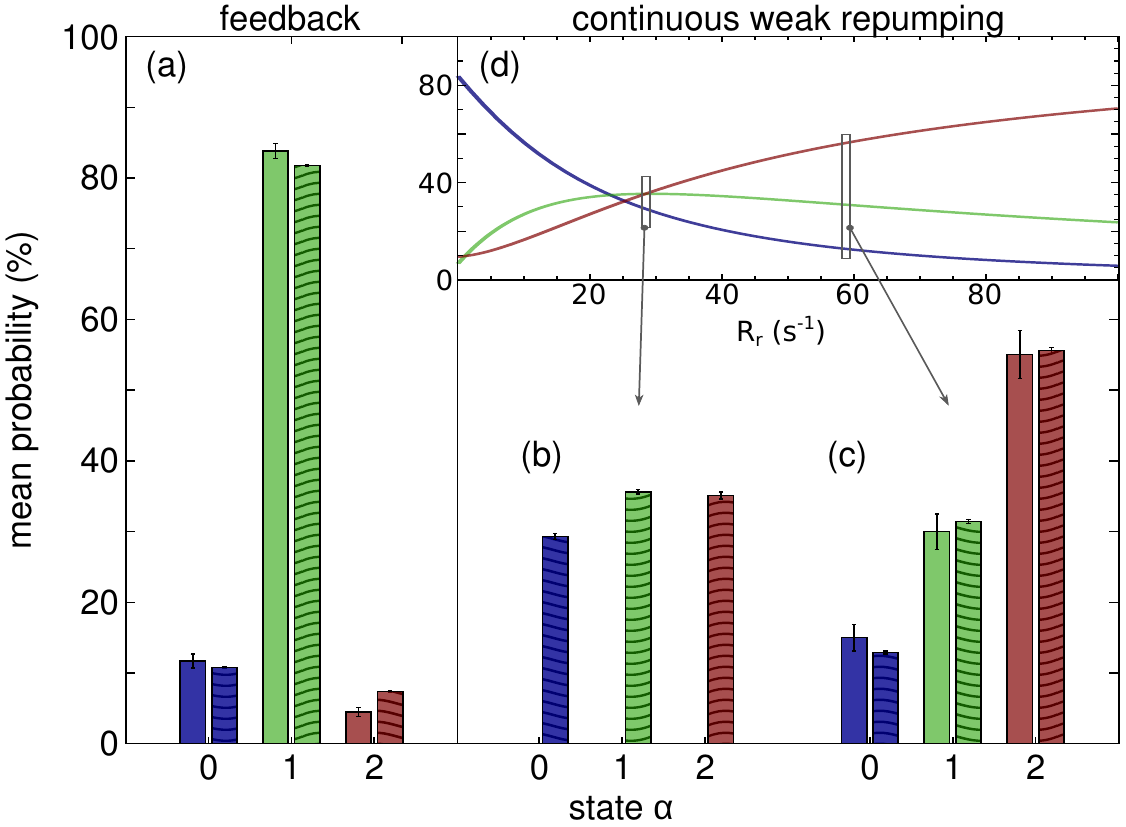}
	\caption{Comparison of mean probabilities for the case of feedback (a) and continuous weak repumping experiments for measured (c) and optimal repumping rate (b). The shaded bars indicate simulated results. (d) Analytical solution of the mean probabilities for different repumping rates $R_\text{r}$.}
	\label{fig_histFeedbackOneState}
\end{figure}

The mean probability of the target state $\hat{\rho}_{\alpha=1}$ over many experiments is $\left< p^\text{post}_1(t)\right>_t = 84\%$, see Fig.~\ref{fig_histFeedbackOneState}(a). This is in excellent agreement with a Monte-Carlo simulation of the feedback based on the measured rates $R_{21}$, $R_{10}$ and the measured photon count rate histograms $p(n(t)|\alpha)$. The scheme is also capable of stabilizing the states $\hat{\rho}_{\alpha = 0,2}$, but these states are trivially accessible by optical pumping.

The time constant for the atom-cavity-system to stay in $\hat{\rho}_{\alpha=1}$ is determined to be $\tau = (19\pm2)\,\text{ms} \approx 1/R_{10}$, in full agreement with the prediction by the rate equations: The mean time in state $\alpha=1$ is ultimately limited by inelastic scattering of the probe laser with rate $R_{10}$ yielding a probability of $1-R_{10}\Delta t$ to stay in this state during a time bin $\Delta t$. The effectivity of the feedback loop is characterized by the time constant until the target state is reached which is experimentally given by $1.12\,\text{ms}$ and thus near the theoretical optimum of a single time bin. The slightly larger mean probability of the lowest state $\hat{\rho}_{\alpha=0}$ compared to $\hat{\rho}_{\alpha = 2}$, visible in Fig.~\ref{fig_histFeedbackOneState}.(a), is caused by the depumping due to the probe laser.

Without feedback the highest passively achievable mean probability of the target state $\hat{\rho}_{\alpha=1}$ is $50\%$ for saturating repumping and depumping lasers. In this limit, the pumping lasers dominate the system dynamics and cause very high transition rates. In a more appropriate case of a weak continuous repumping laser ($R_\text{r}=59\,\text{s}^{-1}$) we have experimentally determined the mean probability $\left< p^\text{post}_1(t)\right>_t = 33\%$, see Fig.~\ref{fig_histFeedbackOneState}(c). The solution of Eq.~(\ref{equation_rateEquation}) for traces of $300\,\text{ms}$ length under the same initial condition of $\mathbf{p}(0) = (0,0,1)^\text{T}$ depending on the repumping rate $R_\text{r}$ yields the expected mean probability as a function of $R_\text{r}$, see Fig.~\ref{fig_histFeedbackOneState}(d). Even at an optimal repumping rate, the mean target state probability never exceeds $37\%$, see Fig.~\ref{fig_histFeedbackOneState}(b).

In order to increase the mean target state probability of our feedback scheme we would have to improve the probability $1-R_{10}\Delta t$ to stay in the target state during a time bin. With an enhanced detection efficiency of the transmitted light we would be able to reduce the bin time. A higher single atom cooperativity would permit to further enhance the cavity-atom detuning for reducing $R_{01}$.

%\section*{Conclusion and outlook}
The stabilized mixed state $\hat{\rho}_{\alpha=1}$ is a statistical mixture of the two Bell states $\ket{\Psi^{\pm}} = (\ket{\uparrow\downarrow} \pm \ket{\downarrow\uparrow})/\sqrt{2}$, which are indistinguishable by our projective transmission measurement. However, the $\ket{\Psi^{-}}$ state is the only eigenstate of a successive application of a common $\pi/2$ single qubit rotation of both atoms and the transmission measurement with a transmission level of $\alpha=1$. Any contribution from $\ket{\Psi^{+}}$ will be projected onto the the states $\hat{\rho}_{\alpha=0(2)}$ after sufficient number of repetitions. Thus a future feedback algorithm might utilize this measurement scheme to detect and purify the entangled state $\ket{\Psi^{-}}$ and to restore it in case of $\hat{\rho}_{\alpha=0(2)}$. For this we would create $\ket{\Psi}^{+}$ according to a probabilistic scheme proposed by S\o rensen and M\o lmer~\cite{Sore03} and convert $\ket{\Psi^{+}}$ to $\ket{\Psi^{-}}$ with a differential phase shift between the two atoms, e.g.~by a magnetic field gradient.

%\begin{acknowledgments}
We acknowledge financial support by the EC through AQUTE and CCQED and by the BMBF through QuOReP. R.R.~acknowledges support from the Studienstiftung des deutschen Volkes and R.R.~and M.M.D.~acknowledge support from the Bonn-Cologne Graduate School of Physics and Astronomy.
%\end{acknowledgments}

\bibliography{Paper}

\end{document}